# Single-Layer Fe-Cu Interphase in Ferritic Steels Stabilized by Magnetic Friedel Oscillations

Wen-Qiang Xie[1,2], Jin-Li Cao[3] and Wen-Tong Geng[2 †]

*1 School of Materials Science and Engineering, Hainan University, Haikou 570228, China*
*2 Department of Physics, Zhejiang Normal University, Jinhua 321004, China*
*[3]Reactor Engineering Technology Research Division, China Institute of Atomic Energy, Beijing 102413, China*

*† Corresponding author: Wen-Tong Geng (E-mail: wtgeng@zjnu.edu.cn)*

## Abstract

Copper precipitation is a technique extensively deployed in steel strengthening. Being as tiny as a few nanometers in diameter, the Cu precipitates present a real challenge to experimental techniques in determination of their composition. The late Professor Morris Fine called it a mystery when addressing the discrepancy between the fact of low solubility of Fe in bulk Cu and the remarkable content of Fe in Cu precipitates according to atom probe tomography measurement. With a thorough search using rigorous first-principles density functional theory calculations, we are surprised to find that the interface between Cu precipitate and Fe matrix is neither immiscibly sharp, nor commonly miscible over a range of atomic layers, but rather manifests itself as a single-layer mixed Fe-Cu interphase. Our detailed analysis reveals that spin polarization is a key factor in defining such an interphase. The magnetic Friedel oscillations through spin polarization is the driving force to stabilize it. When the single-layer Fe-Cu interphase is counted in, the content of Fe in Cu precipitates would be significant due to their small size. Our finding is a strong demonstration that quantum mechanical effects such as Friedel oscillations can have explicit consequence also in structural materials. The accurate knowledge of the Fe-Cu interface is crucial for the design of advanced high-strength steels.

## Introduction

Advanced ultra-high strength steels are highly sought after for a broad spectrum of engineering applications[1]. Recently, nanoscale co-precipitation strengthening in steels has gained significant attention, emerging as a pivotal factor in the development of advanced steels that offer exceptional mechanical, welding, and irradiation properties[2-4]. Particularly, the co-precipitation of diverse nanoparticles is more beneficial than single-type nanoparticle dispersion, as it can result in a superior combination of properties due to the synergistic effects of multiple nanoparticles with varying compositions, microstructures, and micromechanical properties[5-9]. On the other hand, Magnetism in alloy steels significantly impacts their mechanical and electronic properties, influencing phase stability and structural integrity[10,11]. Magnetic Friedel oscillations notably affect surface magnetism[12]. Understanding these effects on the co-precipitation process is crucial for designing materials with tailored properties for advanced engineering applications.

In recent developments, the co-precipitation of nanoscale Cu and AlNi particles in high-strength steels has become a focal point[5,7,13-15], presenting a promising strategy to enhance the strength of steels without markedly compromising ductility. Understanding the precipitation interface mechanisms is essential for integrating computational predictions with experimental validation. It is widely recognized that coherent body-centered cubic (bcc) Cu nanoparticles initially nucleate from the supersaturated solid solution. Subsequently, Ni and Al segregate. Diffusion model based Monte Carlo simulations predict the formation of near-pure Cu clusters at the Fe-Cu interface due to the limited solubility of Fe in Cu and the high diffusivity of Cu in the bcc phase[16]. However, experiment shows that even after aging for 128 hours, atom probe tomography (APT) reveals that Fe remains present within the segregated Cu and AlNi precipitates[17]. This result piqued our interest and prompted the question: why does Fe remain within Cu precipitation particles despite its low solubility in Cu?

To address these phenomena, this study's primary objective is to employ *ab initio* calculations to investigate the interface properties of Fe-Cu alloys. This involves detailed computational modeling to predict the behavior of atoms at the Fe-Cu interface, examining the electronic structure and magnetic interactions. By understanding these fundamental properties, we aim to shed light on how the interface structure influences the overall mechanical properties of the alloy. Additionally, we extend our study to investigate the behavior of Fe-AlNi precipitates to provide a comparative analysis. By comparing these two systems, we can gain a broader understanding of precipitation mechanisms and interface behavior, aiding the development of high-strength steels with optimized properties for engineering applications.

## Computational Details

The calculations were performed using density functional theory (DFT) [18,19] via the Vienna *ab initio* Simulation Package (VASP) [20]. The projector augmented-wave (PAW)[21] method was employed with the Perdew-Burke-Ernzerhof (PBE) generalized gradient approximation (GGA) for the exchange-correlation potential[22] . A plane wave cutoff energy of 400 eV was used, achieving energy convergence within $10^{-6}$ eV. Reciprocal space integrations were sampled with a 6×6×1 Gamma-centered Monkhorst-Pack mesh[23]. The formation energy of Fe-Cu and Fe-AlNi interface is calculated as:

$$E_{f_Fe-cu} = E_{Fe-Cu} - E_{cu} - E_{Fe};$$

$$E_{f_Fe-AlNi} = E_{Fe-AlNi} - E_{Fe} - E_{Al} - E_{Ni};$$

Here, the $E_{Fe-Cu}$, $E_{Fe-AlNi}$, $E_{Fe}$, $E_{cu}$, $E_{Al}$ and $E_{Ni}$ indicates the energy of bcc Fe-Cu, Fe-AlNi, Fe, Cu, Al and Ni, respectively. The interlayer binding energy ($\sum E_{\text{Inter}_i}$) is calculated as follow:

$$\sum E_{\text{Inter}_i} = \sum(E_{\text{Fe-Cu_vac}_i} / E_{\text{Fe-Cu_vac}_i} - E_{\text{Fe-Cu}} / E_{\text{Fe-AlNi}});$$

Here, $E_{\text{Inter}_i}$ indicates the interlayer binding energy of specific layer gap (LG_*i*). $E_{\text{Fe-Cu_vac}_i}$ and $E_{\text{Fe-Cu_vac}_i}$ represent the energy of Fe-Cu/Fe-AlNi with 10Å vacuum at $i_{\text{th}}$ layer gap. Possible interface structures were generated using SAGAR[24]. The average magnetic interaction was studied using OpenMX[25], a DFT-based code, along with a post-processing code based on Green's function[26]. The convergence energy criterion for OpenMX was set to $10^{-5}$ Hartree/Bohr. Interface magnetic coupling values were represented by the integrated value ($J_{\text{int}}$) of average magnetic interaction ($J_{\text{ij}}$)[27]. The integrated value $J_{\text{int}}$ up to the given distance is calculated as:

$$J_{\text{int}} = \frac{1}{2}\sum_{0}^{r} J_{\text{ij}}\, S^2;$$

where $J_{\text{ij}}$ is a function of interface atomic pair distance, $S$ is the unit spin vector. The plane-averaged electron density difference in Fe region is calculated as: $\Delta\rho = \rho_{\text{Fe-Cu}} - \rho_{\text{Fe}}$.

## Results and Discussion

Our investigation assumes a three-layered interface where mixing occurs between Fe and Cu atoms, as visualized in Figure 1 (a) (the grey atoms represent the mixing layers). This exploration identified a total of 349 unique structural configurations (Figure 1 (b)). In Figure 1c, we present the most stable structures across a range of Cu concentrations (Figure1 (c) (d), Figure S1). The formation energy is positive, indicating that Fe and Cu atoms prefer to segregate and form a Fe-Cu interface. Interestingly, the analysis reveals a clear preference for structures exhibiting either: 1. Sharp Fe-Cu interface: This scenario corresponds to the standard Fe-Cu interface model (Figure 2(a)), where Fe and Cu atoms remain largely segregated due to the limited solubility of Fe in Cu, as previously discussed[16]. 2. Single mixed Fe-Cu interface (Figure 2(b)): This configuration suggests a well-defined boundary between the Fe and Cu layers, with a narrow region of intermixing. Notably, the energy landscape favors these two interface types over configurations with more extensive mixing across multiple layers.

The results prompt the question: which model is more energetically preferred (Figure 2 (a))? To investigate this, we allowed atoms within the designated boundary regions in Figure 2 (b) to swap positions, and the concentration of Cu is fixed at 50%. All possible structures are shown in Figure S2. Notably, spin-polarized and non-spin-polarized models exhibit opposite trends. The most stable spin-polarized structure, Stru_0 (sharp interface), is the most unstable without spin polarization, while the most unstable spin-polarized structure, Stru_5 (mixed interface), is the most stable without spin polarization. These calculations reveal that interfaces with some degree of Fe-Cu mixing exhibit lower energy states than sharp Fe-Cu interfaces, challenging existing predictions of pure Cu nanoprecipitation. Without considering spin polarization, the results support the previous study[16] of preference for a sharp Fe-Cu interface.

To fully elucidate the origin of this phenomenon, we compared the interlayer binding energy ($E_{\mathrm{Inter}_i}$) between stable mixing and sharp Fe-Cu interfaces. The total energy ($E_{\mathrm{Total}}$) of the system can be divided into layer and interlayer contributions. Table S1 summarizes the total layer energy ($\sum E_{\mathrm{layer}_i}$) of the pure Fe layer, pure Cu layer, and the Fe-Cu mixed layer, both with and without spin polarization. As expected, the sharp interface shows a stronger energy preference ($\sum E_{\mathrm{layer}_i} = -281.12$ eV) compared to the single mixed layer ($\sum E_{\mathrm{layer}_i} = -280.61$ eV), a trend that holds true even without spin polarization. These findings suggest that the energetic advantage of the single mixed layer may be attributed to its $\sum E_{\text{layer_}i}$.

To confirm the role of $E_{\text{Inter_}i}$ and spin polarization in this system, we employed a 10 Å vacuum layer, allowing it to move upward during calculations (Figure S3). Figure 3(a) presents the variation of $E_{\text{Inter_}i}$ with respect to the layer gap index (LG_*i*) for both sharp and mixed interfaces. The introduction of the interface leads to a noticeable drop in $E_{\text{Inter_}i}$, followed by the formation of Friedel-like oscillations. A significant difference between the spin-polarized and non-spin-polarized sharp interfaces is observed at LG_8, the subsequent interlayer gap of the Fe-Cu interface. Without spin polarization, $E_{\text{Inter_}i}$ exhibits a pronounced drop due to the oscillations. In contrast, the spin-polarized sharp interface shows $E_{\text{Inter_}i}$ oscillation. Consequently, $E_{\text{Inter_}i}$ oscillations without spin polarization enhance the stability of the sharp interface, while spin polarization suppresses these oscillations, reducing the sharp interface's stability.

Figure 3(b) further compares the $E_{\text{Inter_}i}$ for sharp and mixed Fe-Cu interfaces with and without spin polarization. The incorporation of a mixed single layer significantly suppresses the energy oscillation behavior, leaving only a slight oscillation in the mixed interface. Tables S1 summarize the total binding energy for each scenario. Notably, in the absence of spin polarization, $E_{\text{Total}}$ of the sharp interface is 1.73 eV lower than that of the mixed interface. However, with spin polarization, $E_{\text{Total}}$ of the sharp interface becomes 1.16 eV higher than that of the mixed interface. These results indicate that a single mixed layer is energetically favorable when spin-polarization effects are considered, due to the suppression of $E_{\text{Inter_}i}$ oscillations. Figure 3(c) presents the layer-dependent magnetic coupling energy for the sharp interface and vacuum, revealing a clear magnetic Friedel oscillation[12]. However, the presence of the interface suppresses this oscillation. This suppression is corroborated by the changes in the magnetic moment of Fe near the Fe-Cu sharp interface, as shown in Figure 3(d), where the magnetic moment of Fe is significantly reduced. This suggests that interlayer binding oscillations may be counterbalanced by magnetic coupling energy. Figure S4(a) indicates higher electron density in the Fe region compared to the Cu side. This density difference may be attributed to the $E_{\text{Inter_}i}$ oscillation. Furthermore, the plane-averaged electron density difference (Figure S4(b)) indicates that spin polarization results in electron density redistribution. Overall, our observations suggest that interlayer binding energy oscillations and magnetic coupling oscillations may counteract each other.

In addition to investigating the Fe-Cu interface, we also examined the Fe-AlNi interface for comparative analysis. Similar to the approach in Figure 2(b), we focused on atomistic exchanges around the two interfaces, with possible structures depicted in Figure S5. Figure S6 presents the energetically favorable configurations with and without spin polarization, revealing a preference for a mixed interface in both cases, albeit with different specific configurations. To elucidate the differences between the Fe-AlNi and Fe-Cu interfaces, we analyzed the $E_{\text{Inter_}i}$ variation for Fe-AlNi, shown in Figure 4(a). Spin polarization suppresses the $E_{\text{Inter_}i}$ at the Ni end but has a milder effect at the Al end. As shown in Table S2, without spin polarization, the mixed interface's $\sum E_{\text{layer_}i}$ is 3.79 eV lower and its $\sum E_{\text{Inter_}i}$ is 3.71 eV higher than the sharp interface, making the $E_{\text{Total}}$ only 0.08 eV lower, slightly favoring the mixed interface. With spin polarization, the $E_{\text{Total}}$ of the mixed interface is 6.77 eV lower, further favoring the mixed interface. This result implies a stronger intralayer interaction in Fe-AlNi interface. Figure 4(b) illustrates the impact of spin polarization, which significantly affects the magnetic moment of Fe near the Ni end while having a smaller effect near the Al end. This accounts for the disappearance of the oscillation cone near the Ni end, while it remains near the Al end, as shown in Figure 4(a). Figure S7(a) and (b) present the Electron density contour map and the plane-averaged electron density difference of Fe atoms. Figure S7(c) shows the interlayer binding energy of the mixed interface Spin polarization appears to have minimal impact on the binding energy oscillation within the Fe region, further confirming that the differences in specific mixed interface configurations are attributed to intralayer coupling effects. Overall, in the Fe-AlNi system, stronger intralayer interactions lead to a preference for the mixed interface regardless of spin polarization. Additionally, the suppression of $E_{\text{Inter_}i}$ oscillation further supports this preference.

## Conclusion

We have investigated 349 distinct structural configurations of the Fe-Cu interface, confirming that the interface can manifest either as a single-layer mixed interface or as a sharp Fe-Cu boundary. Our detailed analysis indicates that spin polarization plays a crucial role in determining these two interface structures. we observe interlayer binding energy oscillations that favor the formation of a sharp interface. However, the inclusion of spin polarization introduces magnetic Friedel oscillations. These two types of oscillations may counteract each other, resulting in the suppression of the overall oscillatory behavior and thus favoring the formation of a mixed interface. In the Fe-AlNi system, the stronger intralayer interaction results in a preference for the mixed interface, regardless of spin polarization. Our findings are significant for advancing the understanding of interface behavior in these alloy systems, providing insights that are crucial for the design and development of advanced high-strength steels.

## Acknowledgments

This research was supported by the natural science fund projects of Hainan Province-Youth fund projects (Grants No. 124QN178).

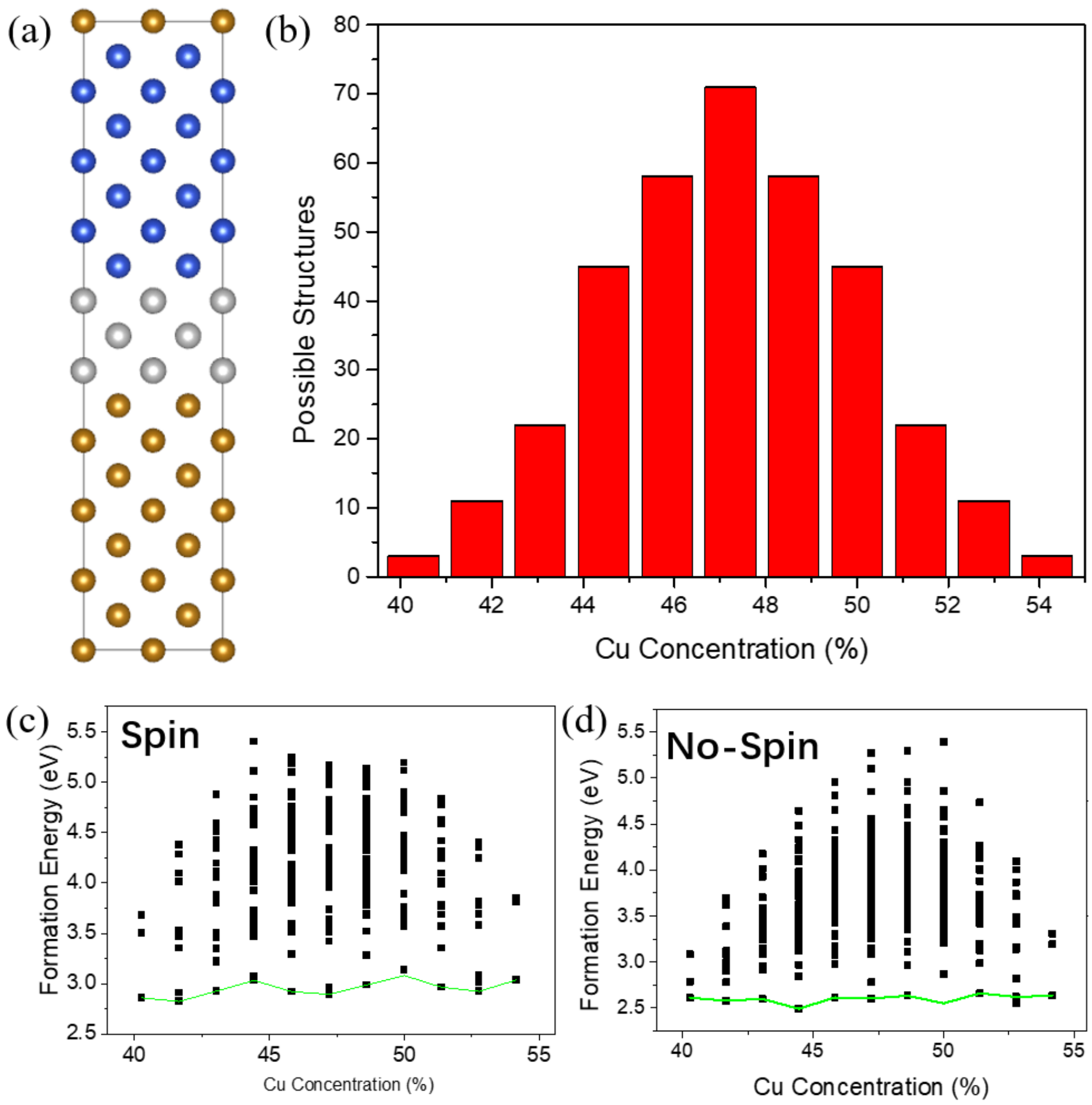

Figure 1: (a) Schematic representation of the Fe-Cu interface, where yellow spheres represent Fe atoms, blue spheres represent Cu atoms. The grey spheres represent the mixed layers, where positions can be occupied by either Fe or Cu atoms. (b) Nonequivalent structures with varying Cu concentrations. Formation energy as a function of Cu concentration (c) with spin-polarization considerations and (d) without spin-polarization considerations.

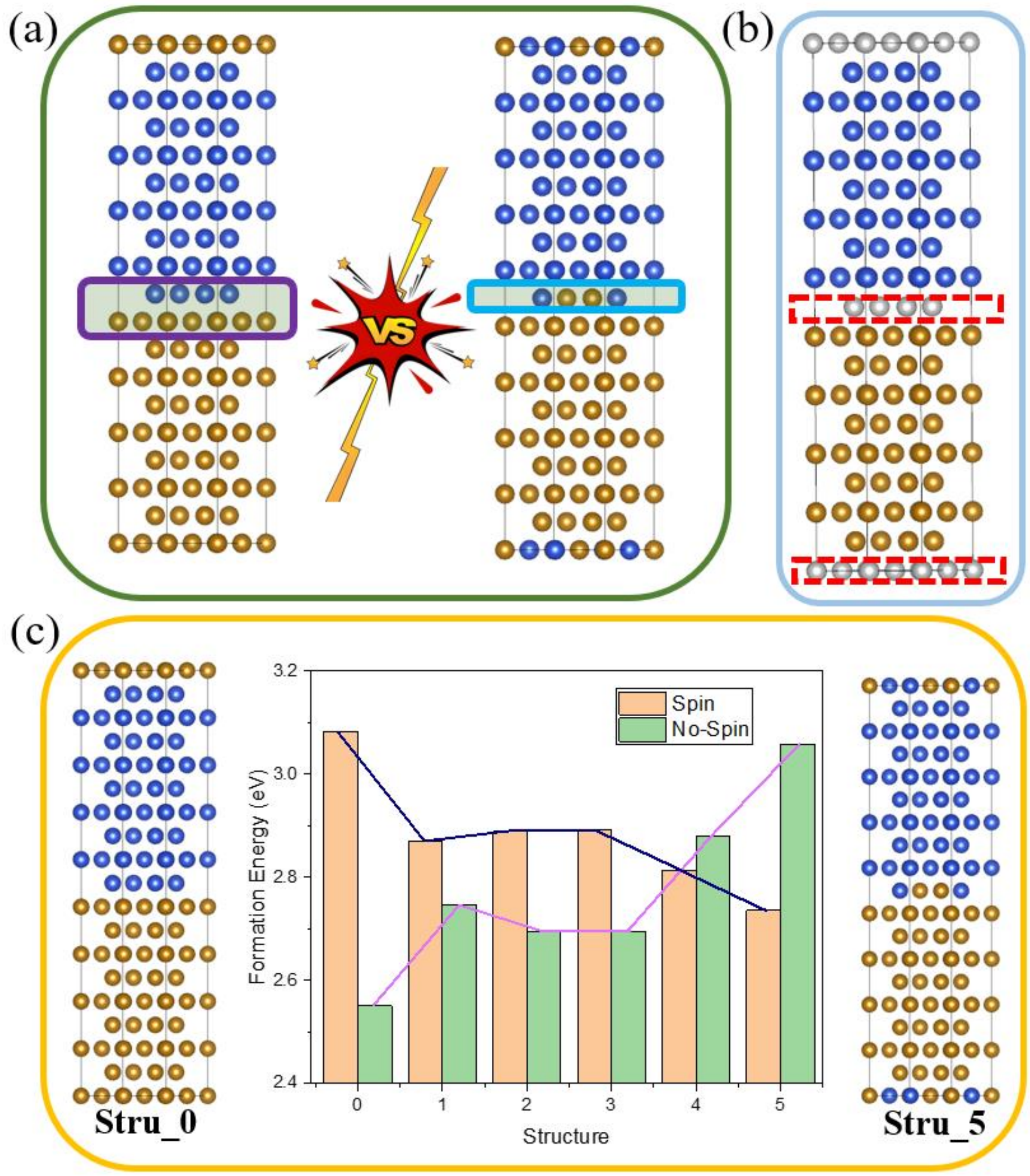

Figure 2: (a) Comparison between sharp and mixed interfaces. (b) Models used in calculation, where grey spheres indicate positions that can be occupied by either Fe or Cu atoms, the concentration of Cu is fixed at 50%. (c) Formation energies of different models: Stru_0 represents a sharp interface, while Stru_1 to Stru_5 represent mixed interfaces.

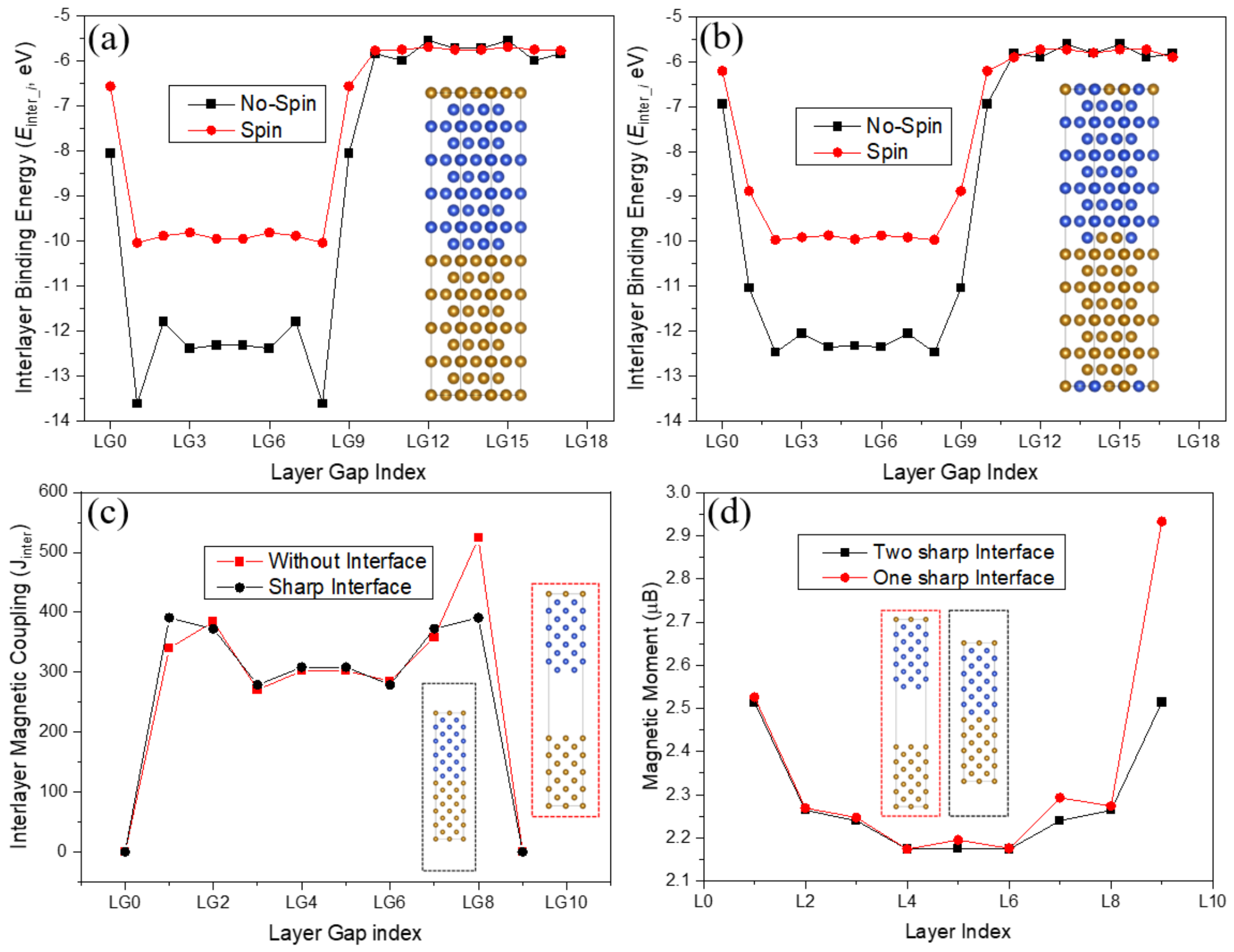

Figure 3: Interlayer Binding energy of the (a) Fe-Cu sharp interface and (b) mixing interface with and without spin-polarization. Dependence of (c) interlayer magnetic coupling and (d) magnetic moment on layer gap and layer index.

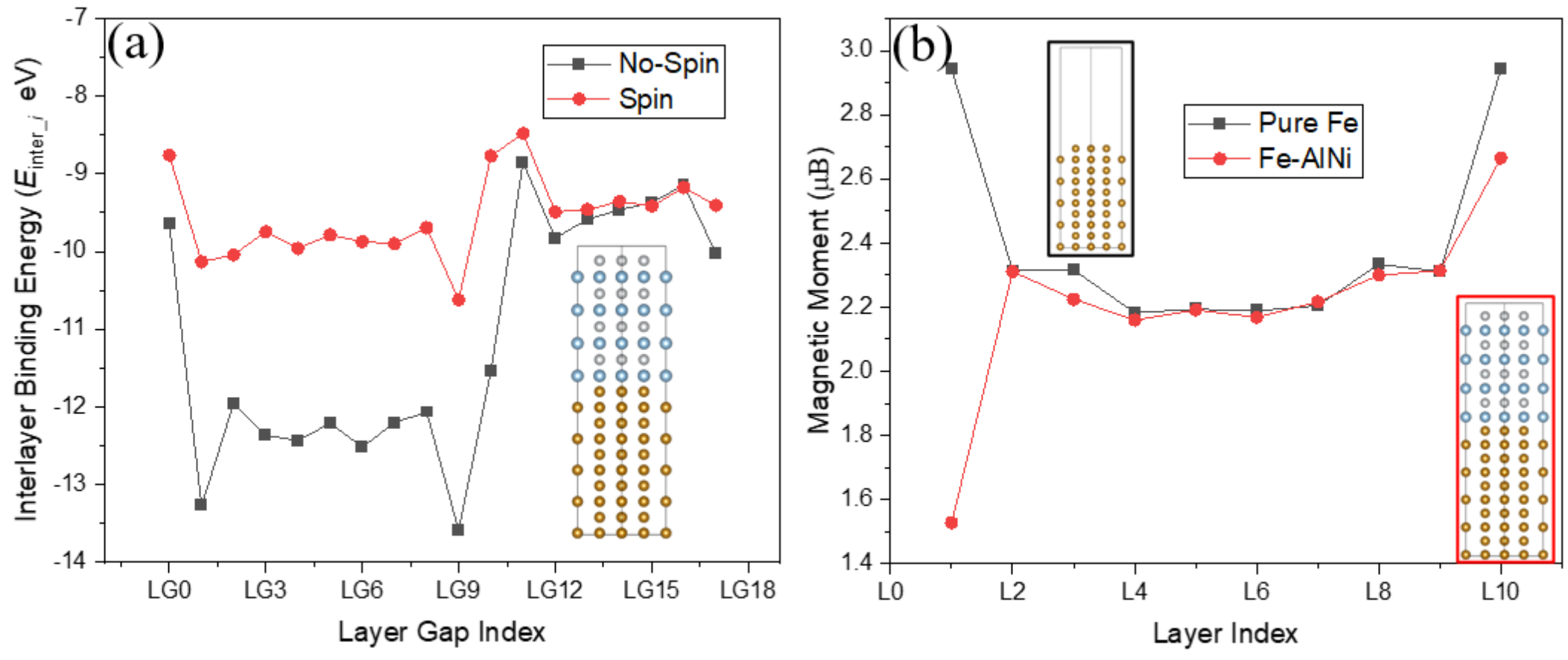

Figure 4: Interlayer Binding energy of the (a) Fe-AlNi sharp interface with and without spin-polarization. (b) Magnetic moment of pure Fe and Fe-AlNi sharp interface.

# Supplementary

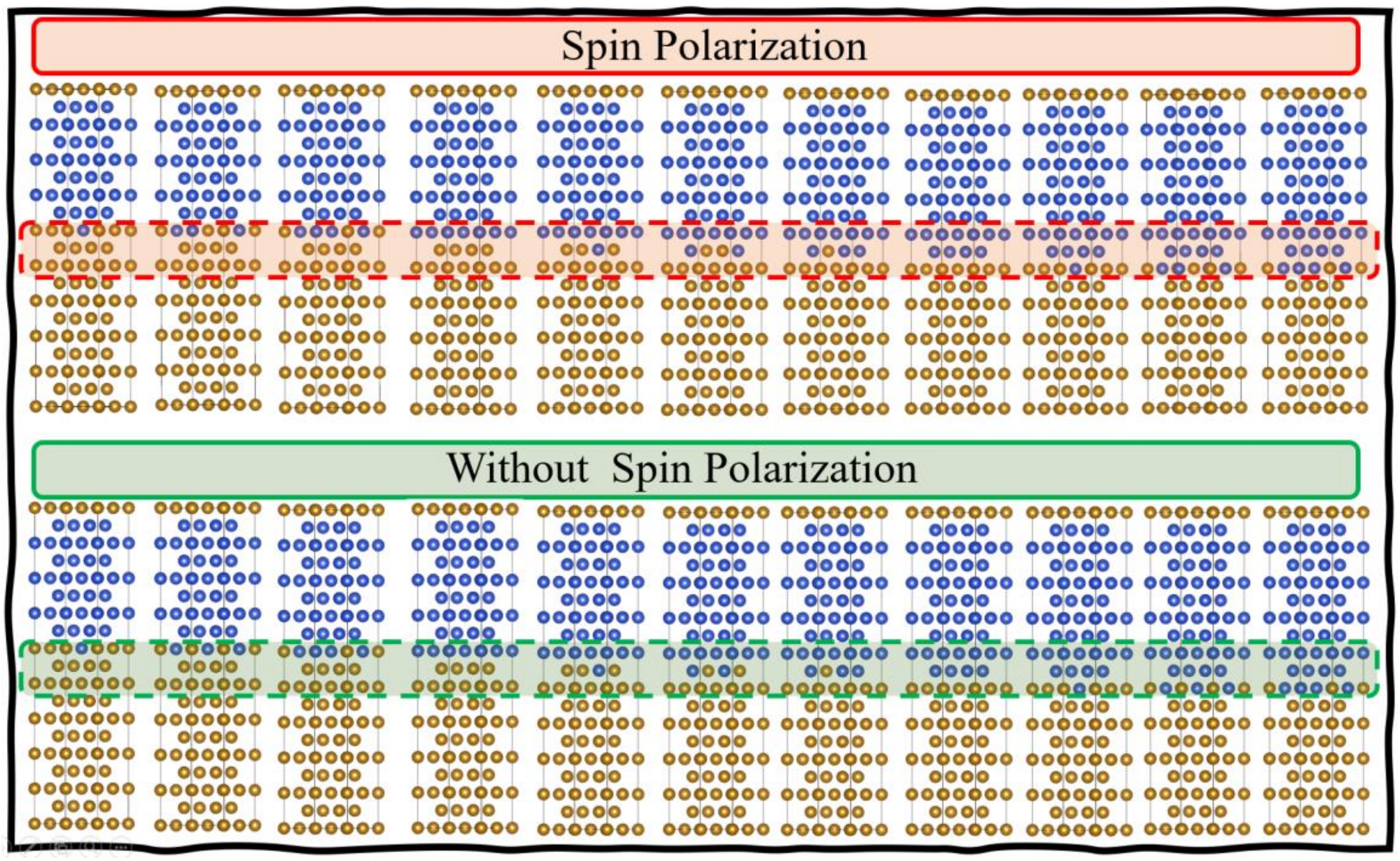

Figure S1: The energy favorable structures with and without spin polarization.

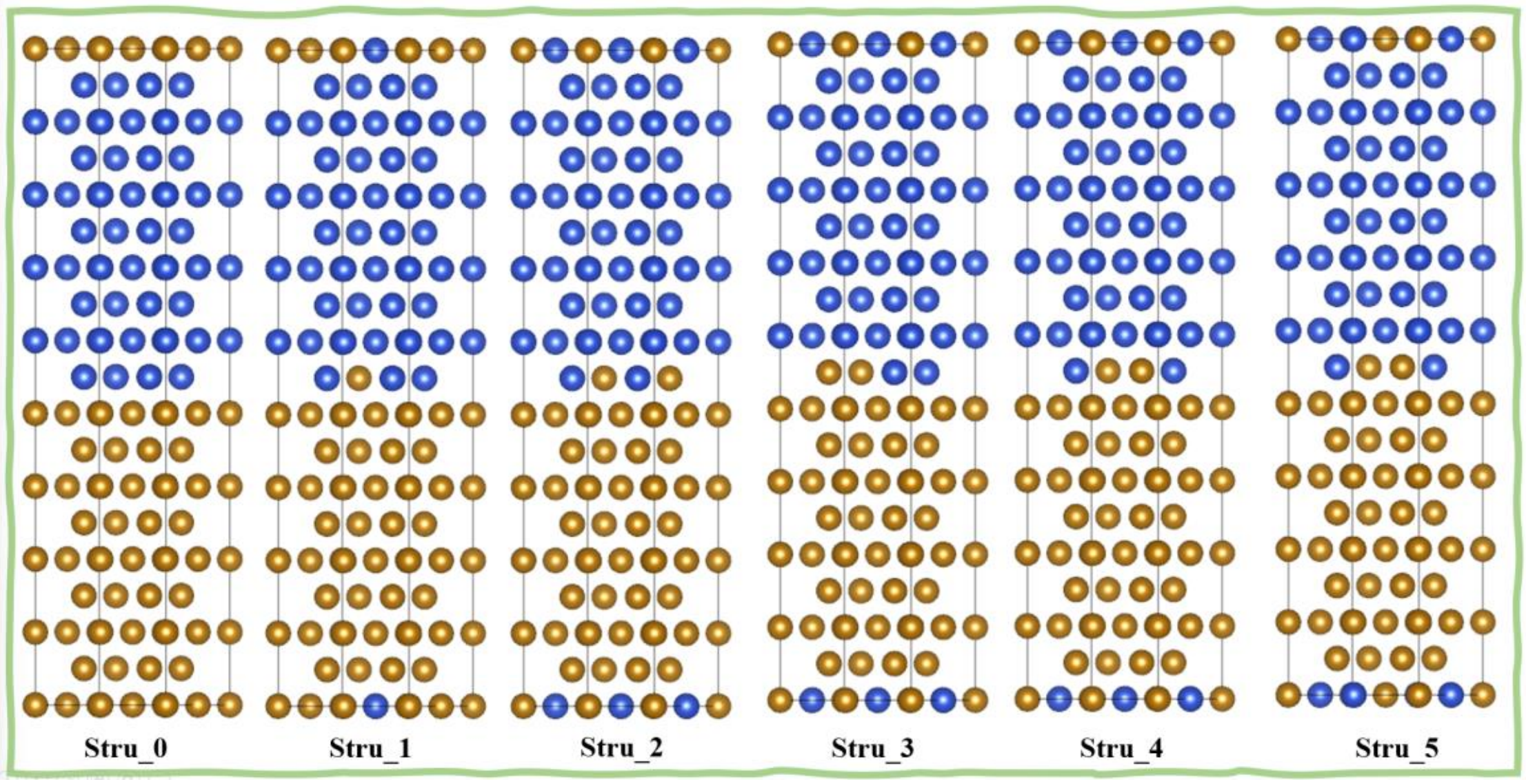

Figure S2: All possible structures if atoms within the designated boundary regions are allowed to swap positions. Here, the concentration of Cu is fixed at 50%.

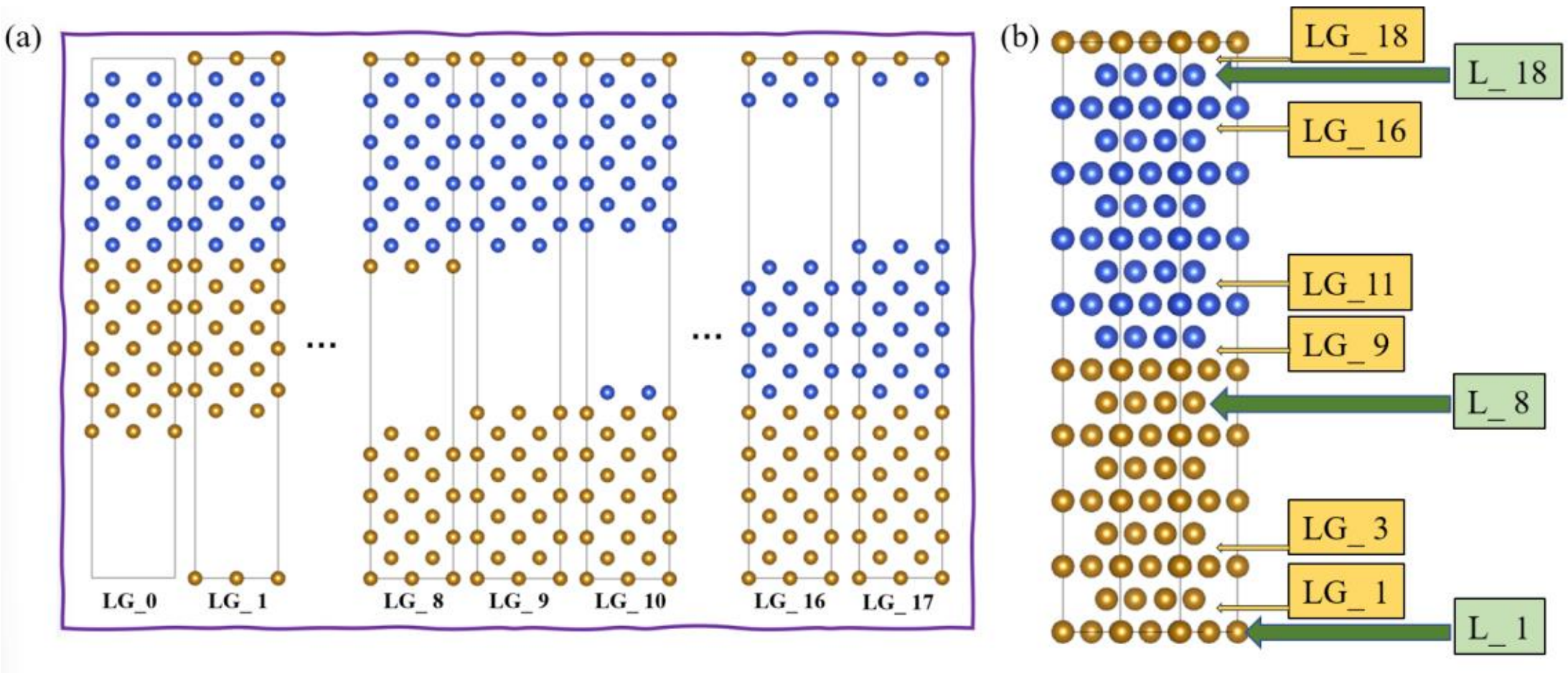

Figure S3: (a) Illustration of the models used to calculate the interlayer binding energy. (b) The illustration of layer index (L_i) and layer gap indes (LG_i).

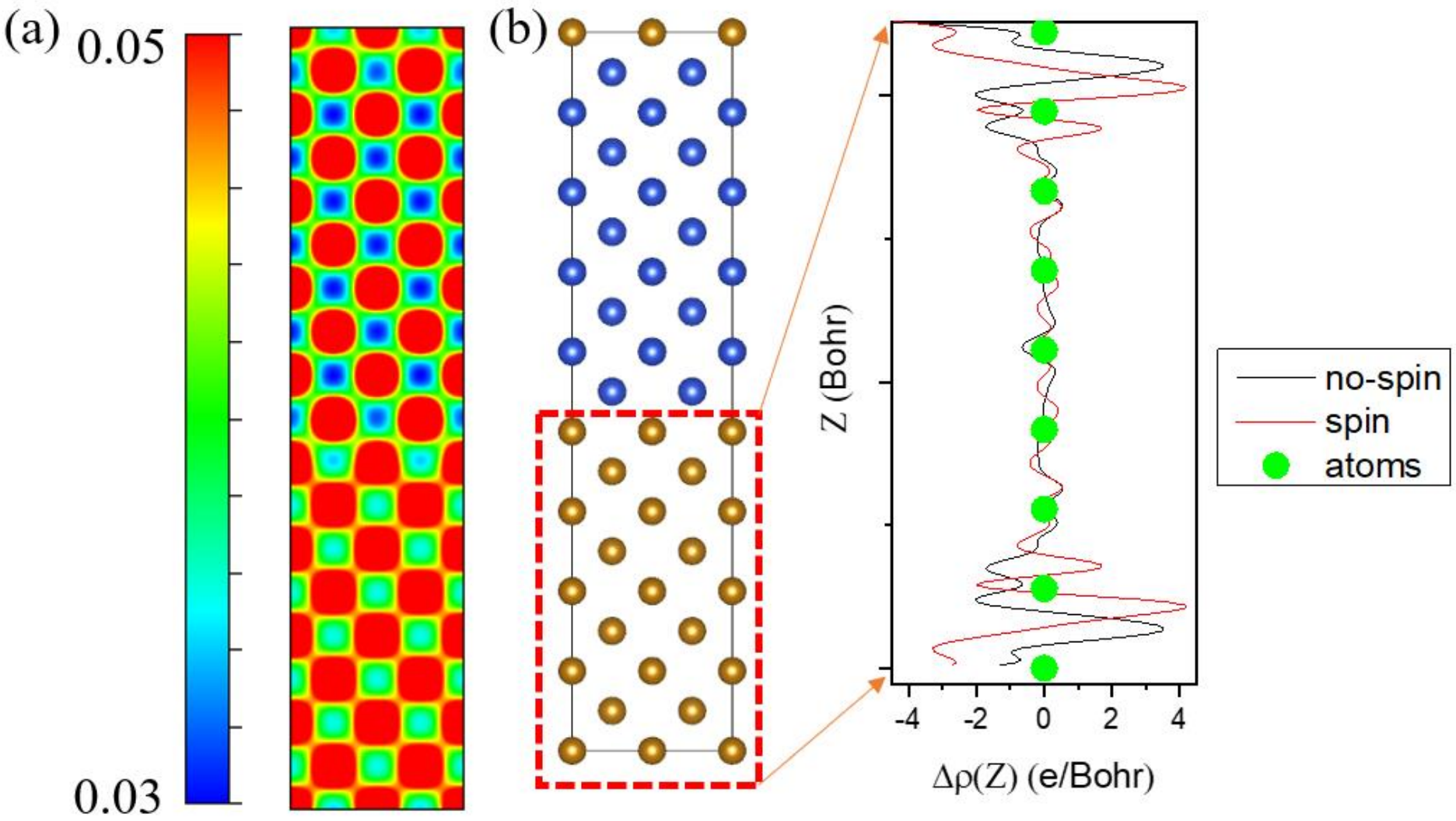

Figure S4: (a) Electron density contour map from 0.03 to 0.05 e/Å$^3$. (b) The plane-averaged electron density difference in Fe region.

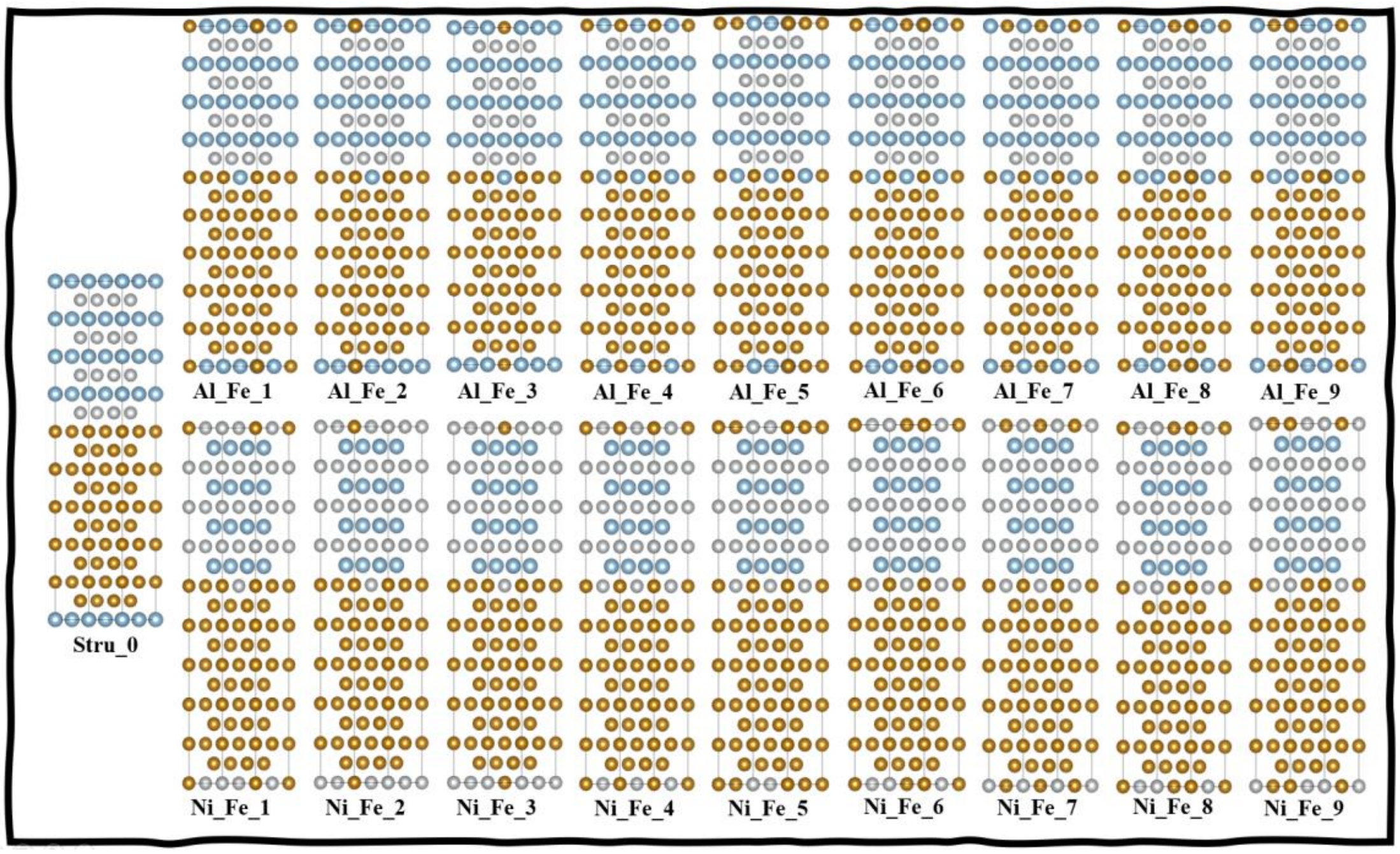

Figure S5: The possible configurations where the Al and Ni concentration is 22.22%.

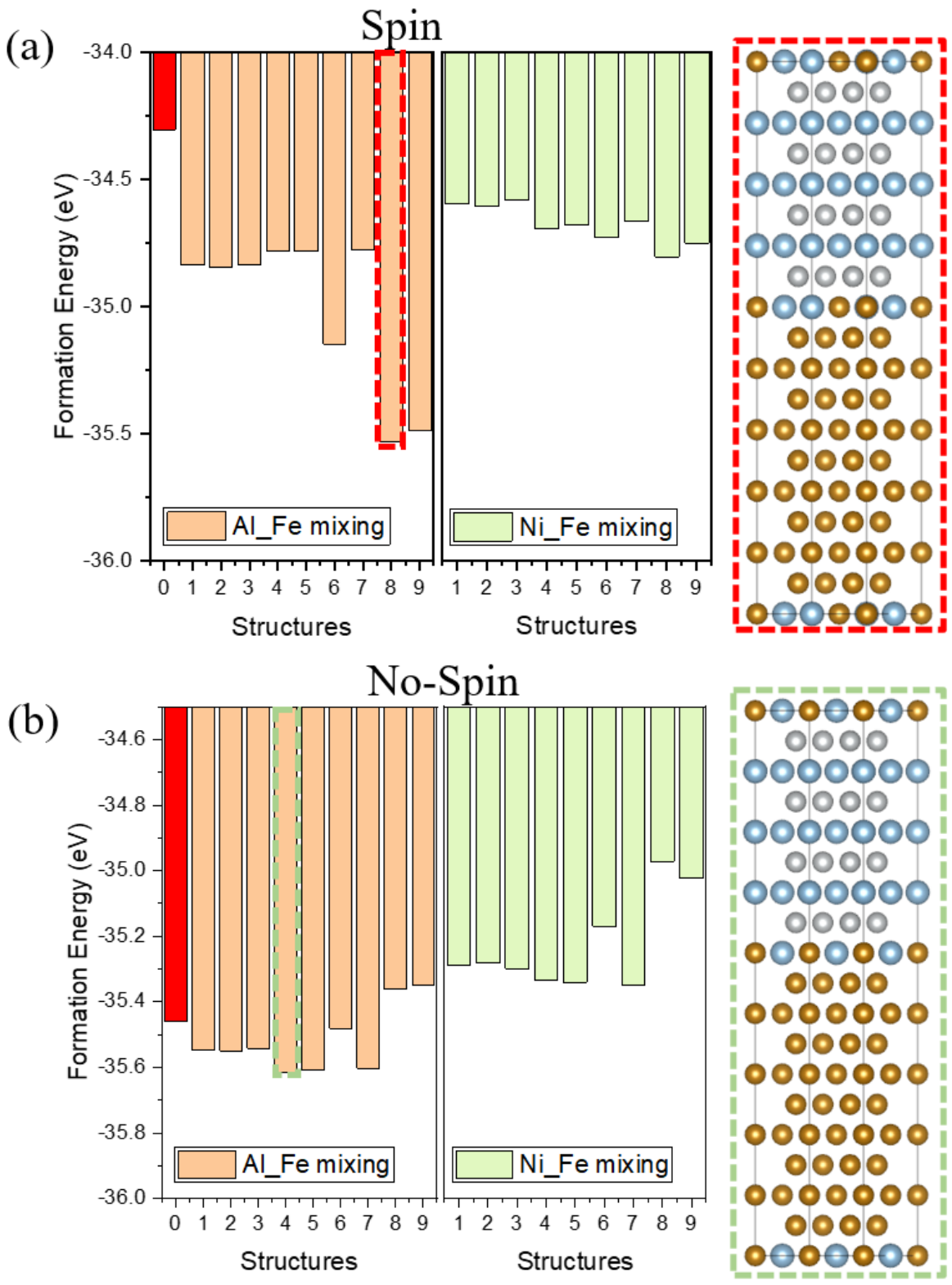

Figure S6: The energy favorable structures in Figure S4 with (a) and without (b) spin-polarization.

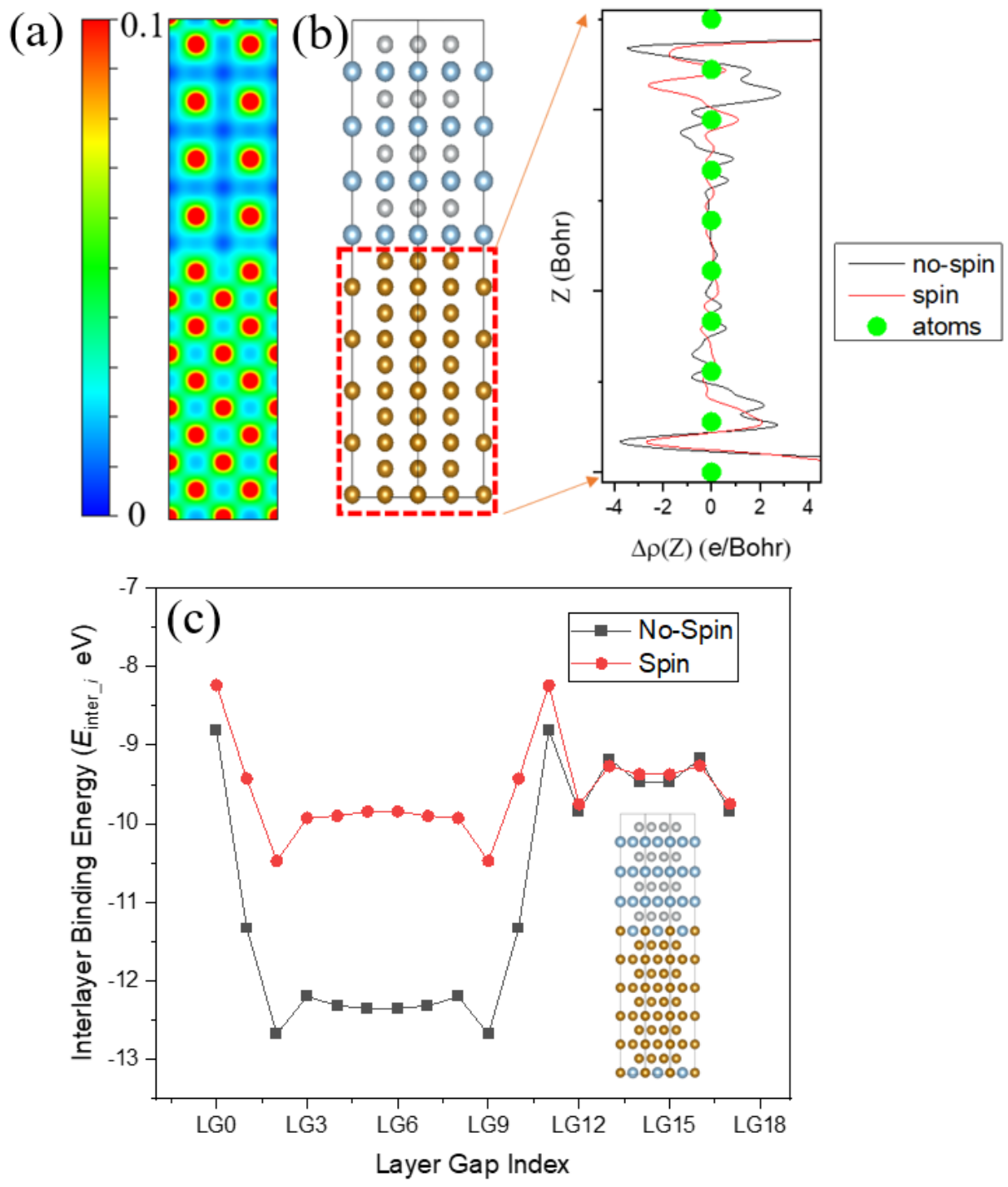

Figure S7: (a) Electron density contour map from 0 to 0.1 e/Å$^3$. (b) The plane-averaged electron density difference in Fe region. (c) Interlayer binding energy of the Fe-AlNi mixing interface.

Table S1: The interlayer binging ($\sum E_{\text{Inter_}i}$), layer ($\sum E_{\text{layer_}i}$) and total energy ($E_{\text{Total}}$) of Fe-Cu sharp and single mixing interface.

| | **No-Spin** | | **Spin** | |
|---|---|---|---|---|
| interface | sharp | mixing | sharp | mixing |
| $\sum E_{\text{Inter_}i}$ (eV) | -162.58 | -162.53 | -138.48 | -140.15 |
| $\sum E_{\text{layer_}i}$ (eV) | -225.59 | -223.91 | -281.12 | -280.61 |
| $E_{\text{Total}}$ (eV) | -388.17 | -386.44 | -419.60 | -420.76 |

*$E_{\text{Total}} = \sum E_{\text{Inter_}i} + \sum E_{\text{layer_}i}$;

Table S2: The $\sum E_{\text{Inter_}i}$, $\sum E_{\text{layer_}i}$ and $E_{\text{Total}}$ of Fe-AlNi sharp and single mixing interface.

| | **No-Spin** | | **Spin** | |
|---|---|---|---|---|
| interface | sharp | mixing | sharp | mixing |
| $\sum E_{\text{Inter_}i}$ (eV) | -200.05 | -196.34 | -172.05 | -172.40 |
| $\sum E_{\text{layer_}i}$ (eV) | -253.70 | -257.49 | -317.70 | -324.11 |
| $E_{\text{Total}}$ (eV) | -453.75 | -453.83 | -489.74 | -496.51 |

*$E_{\text{Total}} = \sum E_{\text{Inter_}i} + \sum E_{\text{layer_}i}$;